\documentclass[useAMS,galley]{mn2e}
\usepackage{graphicx}
\newcommand{\ltsima}{$\; \buildrel < \over \sim \;$}
\newcommand{\simlt}{\lower.5ex\hbox{\ltsima}}
\newcommand{\gtsima}{$\; \buildrel > \over \sim \;$}
\newcommand{\simgt}{\lower.5ex\hbox{\gtsima}}

\newcommand{\cgs}{ ${\rm erg~cm}^{-2}~{\rm s}^{-1}$} 
\newcommand{\lum}{\rm erg~s$^{-1}$}

\def\lesssim{\mathrel{\hbox{\rlap{\hbox{\lower4pt\hbox{$\sim$}}}\hbox{$<$}}}}
\def\gtrsim{\mathrel{\hbox{\rlap{\hbox{\lower4pt\hbox{$\sim$}}}\hbox{$>$}}}}

\def\arcsec{\hbox{$^{\prime\prime}$}}

\def\ab1450{$AB_{1450(1+z)}$}

\def\xray{\hbox{X-ray}}
\def\feka{Fe\ K$\alpha$}
\def\oviii{\hbox{O\ {\sc viii}}}
\def\ovii{\hbox{O\ {\sc vii}}}
\def\fs{$F_{\rm 0.5-2\ keV}$}
\def\fh{$F_{\rm 2-10\ keV}$}

\def\today{\ifcase\month\or January\or February\or March\or April\or May\or
      June\or July\or August\or September\or October\or November\or December\fi
      \space\number\day, \number\year}
\def\asca{{\it ASCA\/}}
\def\chandra{{\it Chandra\/}}

\def\fuse{{\it {\it FUSE}\/}}

\def\heao1{{\it HEAO-1\/}}
\def\hst{{\it {\it HST}\/}}

\def\rxte{{\it RXTE\/}}
\def\sax{{\it BeppoSAX\/}}

\def\xmm{{XMM-{\it Newton\/}}}

\title[\xmm\ OBSERVATIONS OF ARAKELIAN~564]{Arakelian~564: An \xmm\ view}
\author[C. Vignali et al.]
{
Cristian Vignali,$^{1,2}$\thanks{E-mail: chris@nellie.astro.psu.edu (CV); niel@astro.psu.edu (WNB); 
bol@xray.mpe.mpg.de (TB); acf@ast.cam.ac.uk (ACF); sav2@star.le.ac.uk (SV).}
W.~N. Brandt,$^{1}$\footnotemark[1]
Th. Boller,$^{3}$\footnotemark[1] 
A.~C. Fabian$^{4}$\footnotemark[1] and 
\newauthor 
Simon Vaughan$^{4,5}$\footnotemark[1] \\ \\ 
$^{1}$ Department of Astronomy and Astrophysics, The Pennsylvania State University, 
525 Davey Laboratory, University Park, PA 16802, USA \\
$^{2}$ INAF -- Osservatorio Astronomico di Bologna, Via Ranzani 1, 40127 Bologna, Italy \\
$^{3}$ Max-Planck-Institut f\"ur extraterrestrische Physik, 
Postfach 1312, 85741 Garching, Germany \\
$^{4}$ Institute of Astronomy, Madingley Road, Cambridge, CB3~0HA \\
$^{5}$ Department of Physics and Astronomy, University of Leicester, 
Leicester, LE1 7RH \\
}

\begin{document}

\date{Accepted 2003 ???. Received 2003 ???; in original form 2003 ???}

\pagerange{\pageref{firstpage}--\pageref{lastpage}} \pubyear{2003}

\maketitle

\label{firstpage}

\begin{abstract}
We report on two \xmm\ observations of the bright narrow-line Seyfert~1 galaxy Ark~564 taken 
one year apart (2000 June and 2001 June). 
The \hbox{0.6--10~keV} continuum is well described by a soft blackbody component 
(\hbox{$kT\approx$~140--150~eV}) plus a steep power law (\hbox{$\Gamma\approx$~2.50--2.55}). 
No significant spectral changes are observed between the two observations, 
although the \xray\ flux in the second observation is $\approx$~40--50 per cent lower. 
In both observations we detect a significant absorption edge at a rest-frame energy of 
\hbox{$\approx$~0.73~keV}, corresponding to \ovii. 
The presence of the absorption feature is confirmed by a 
simultaneous \chandra\ grating observation in 2000 June, although the best-fitting edge threshold 
is at a slightly lower energy in the \chandra\ data, possibly because of a 
different parameterisation of the underlying \xray\ continuum. 
%
We find tentative evidence for a broad iron emission line in the 2000 June observation. 
The results from an analysis of the power spectral density (PSD) function are also presented. 
The present \xmm\ data support the idea that the PSD shows two breaks, although the location of the 
high-frequency break requires further constraints. 
\end{abstract}


\section{Introduction}

Narrow-Line Seyfert~1 galaxies (NLS1s) are a subclass of active galactic nuclei (AGNs) 
defined by their exceptionally narrow optical permitted emission lines 
\hbox{(H$\beta$ FWHM $<2000$~km~s$^{-1}$)} and their generally high \hbox{Fe\ II/H$\beta$} ratios 
(e.g., Osterbrock \& Pogge 1985; Goodrich 1989), 
in comparison with ``normal'' broad-line Seyfert~1 galaxies (BLS1s). 
NLS1s often show extreme AGN properties; their ultraviolet (UV)/optical emission lines 
place them at one extreme of the Boroson \& Green (1992) 
primary eigenvector. In the \xray\ regime, NLS1s are characterized by rapid and large-amplitude 
variability (e.g., Boller, Brandt \& Fink 1996; Boller et al. 1997); 
the normalised excess variance of variability 
is systematically larger in NLS1s than in BLS1s 
(e.g., Leighly 1999a; Turner et al. 1999; Cancelliere \& Comastri 2002), 
despite both classes having similar \xray\ luminosity distributions. 
The \xray\ spectra of NLS1s below \hbox{$\approx$~1--1.5~keV} often show a prominent soft excess, 
modeled either by a steep power law or a thermal component (e.g., Brandt 1999; Comastri 2000). 
Moreover, their hard ($>2$~keV) \xray\ spectra are generally characterized by 
steeper power-law photon indices (e.g., Brandt, Mathur \& Elvis 1997; Comastri 2000) 
than those of BLS1s.   
A likely explanation for the different properties of NLS1s is that they 
have relatively low masses for their central black holes and high accretion rates 
(e.g., Czerny et al. 2001; Boroson 2002; Wang \& Netzer 2003). 
Smaller black hole masses can naturally explain both the narrowness of the optical emission lines, which are 
generated in gas that has smaller Keplerian velocities, and the extreme \xray\ variability, since the primary emission 
would originate in a smaller region around the central engine. Soft photons from the accretion disc may 
Compton cool electrons in the corona and cause the steep observed photon indices 
(Pounds, Done \& Osborne 1995; Haardt, Maraschi \& Ghisellini 1997). 
In the case of high accretion rates, the surface of the disc is expected to be ionised 
(e.g., Matt, Fabian \& Ross 1993; Ballantyne, Iwasawa \& Fabian 2001). 
The disc will thus produce ionised \feka\ features, as have apparently been observed from some NLS1s 
(e.g., Comastri et al. 1998; 2001; Vaughan et al. 1999b; Leighly et al. 1999b; Turner et al. 2001a). 

Arakelian~564 (hereafter Ark~564; $z=0.0247$) is one of the \xray\ brightest NLS1s 
(e.g., Brandt et al. 1994; Vaughan et al. 1999a,b). 
In 2000 June it was the subject of an intensive multiwavelength monitoring campaign 
that included simultaneous observations with \asca\ (Turner et al. 2001b, hereafter T01; 
Edelson et al. 2002), 
\rxte\ (Pounds et al. 2001), \chandra\ (Matsumoto, Leighly \& Marshall 2001; Marshall 2002), 
\xmm\ (this paper), \fuse\ (Romano et al. 2002), and \hst\ (Collier et al. 2001; Crenshaw et al. 2002). 
Ark~564 represents a good target for \xmm: the low-energy coverage allows accurate modeling of the 
soft excess, while the relatively large effective area at high energies allows studies of the 
previously revealed \feka\ emission line (e.g., Comastri et al. 2001). 
Furthermore, the large \xmm\ count rate allows the best possible studies of rapid variability.

\section{Data reduction}

Ark~564 was observed by \xmm\ (Jansen et al. 2001) 
on 2000 June 17 (rev.~96) and 2001 June 9 
(rev.~275).\footnote{Ark~564 was observed twice because of operational difficulties
that prevented the entire accepted exposure from being obtained in one observation.}
%
\begin{table}
\centering
\caption{Ark~564: \xmm\ observations and source statistics.}
\begin{center}
\begin{tabular}{lccccc}                
\hline
       & RGS1 & RGS2 & pn & MOS1 & MOS2 \\
\hline
\multicolumn{6}{c}{Obs. ID=0006810201; Rev=96; Obs. Date=2000 June 17} \\
\hline
Exp. Time (ks) & 5.59  & 5.43  & \dotfill & \dotfill & \dotfill \\
Source Counts  & 15929 & 14111 & \dotfill & \dotfill & \dotfill \\
\hline
\multicolumn{6}{c}{Obs. ID=0006810101; Rev=96; Obs. Date=2000 June 17} \\
\hline
Exp. Time (ks) & 11.68  & 11.47  & 6.10   & 6.30  & 6.53  \\
Source Counts  & 25449  & 23401  & 307660 & 72359 & 80861 \\
\hline
\multicolumn{6}{c}{Obs. ID=0006810401; Rev=275; Obs. Date=2001 June 9} \\
\hline
Exp. Time (ks) & 6.94  & 6.30  & \dotfill & \dotfill & \dotfill \\
Source Counts  & 9002  & 8871  & \dotfill & \dotfill & \dotfill \\
\hline
\multicolumn{6}{c}{Obs. ID=0006810301; Rev=275; Obs. Date=2001 June 9} \\
\hline
Exp. Time (ks) & 3.93 & 4.10 & 3.68   & 5.87  & 5.84  \\
Source Counts  & 4463 & 4546 & 111450 & 43454 & 43719 \\
\hline
\end{tabular}
\begin{minipage}{85mm}
The reported exposure times refer to the ``cleaned'' intervals 
(after removing the periods of flaring background) in the \hbox{0.3--2~keV} 
band for the RGS and in the \hbox{0.3--10~keV} band for the 
EPIC pn and MOS. 
\end{minipage}
\end{center}
\label{tab:tab1}
\end{table}
%

At the beginnings of both observations the European Photon Imaging Camera (EPIC) 
instruments [pn (Str\"{u}der et al. 2001) and MOS (Turner et al. 2001)] were closed, while the 
Reflection Grating Spectrometer (RGS; den Herder et al. 2001) and the Optical Monitor (OM; Mason et al. 
2001) were operated. 
The EPIC cameras were operated in ``Small-Window'' mode to mitigate ``pile-up'' of photons  
from the source; EPIC data were acquired using the medium optical blocking filter. 
The data were reduced with the latest version of the \xmm\ Science Analysis Software 
({\sc sas}; version 5.4.1) using the latest calibration products.  
The data were filtered to avoid background flares 
(due to $\approx$~100~keV protons; e.g., De Luca \& Molendi 2002) which affect large intervals 
($\approx$~50\%) of our observations. 
For the EPIC data, the tasks {\sc epproc} and {\sc emproc} were used to generate valid photon list.  
Both single-pixel and double-pixel events (patterns 0--4) were used 
when extracting the pn counts, while patterns 0--12 were used for the MOS. 
Source counts were extracted in the \hbox{0.3--12~keV} band 
from circular regions of radius 90\arcsec\ (pn) and 50\arcsec\ (MOS); background counts 
were extracted from off-source regions of radius 45\arcsec\ (pn) and 150\arcsec\ (MOS). 
The statistics are clearly dominated by the source counts in any chosen EPIC sub-band. 
Response functions for spectral fitting to the EPIC data were generated using the tasks {\sc rmfgen} and {\sc arfgen}. 
First-order RGS spectra were extracted using the {\sc rgsproc} task, which also produced the appropriate response matrices. 
A summary of the \xmm\ observations of Ark~564 along with the source counts are  
presented in Table~1. 
Note the large numbers of counts obtained with the present \xmm\ observations of Ark~564; 
for comparison, 
each \asca\ Solid-state Imaging Spectrometer gathered \hbox{$\approx$~2$\times10^{6}$}  
source counts 
over the whole 35-day observation (T01) 
and $\approx$~12000 in the same time interval as the \xmm\ observation (see $\S$3.1). 

\section{Spectral analysis}

The source counts were grouped into spectra such that each spectral bin contained at least 
20 counts to allow $\chi^{2}$ fitting. 
\xray\ spectra were fitted using the {\sc xspec} package (version 11.2; Arnaud 1996). 
The quoted errors on derived model parameters correspond to the 90 per cent confidence level 
for one interesting parameter (i.e., $\Delta\chi^{2}=2.71$; Avni 1976) unless otherwise stated. 
All spectral fits include absorption due to the line-of-sight Galactic column density of 
\hbox{$N_{\rm H}=6.4\times10^{20}$~cm$^{-2}$} (Dickey \& Lockman 1990). 
Hereafter we adopt \hbox{$H_{0}$=70~km~s$^{-1}$~Mpc$^{-1}$} in a $\Lambda$-cosmology 
with \hbox{$\Omega_{\rm M}$=0.3} and \hbox{$\Omega_{\Lambda}$=0.7}. 

At first the MOS1 and MOS2 spectra were fitted separately 
to check for cross-calibration uncertainties; 
these have been found to be $\approx$~3\% (Kirsch 2003). 
We found generally good agreement between MOS1 and MOS2, comparable with the value quoted above. 
Therefore in the following we will present the spectral results of both cameras referred to as MOS. 

Cross-calibration uncertainties between pn and MOS are generally at most $\approx$~13\% 
(Kirsch 2003). However, the pn and MOS spectra of Ark~564 in both observations 
clearly differ at energies below \hbox{$\approx$~0.6~keV}; 
this is probably due to residual calibration problems in the ``Small-Window'' mode. 
A similar difference in the pn/MOS behaviour at low energies has been found for 
PG~1211$+$143 (see Fig.~3 of Pounds et al. 2003). 
Therefore, in the following the pn and MOS data will be fitted together with the same model 
in the \hbox{0.6--10~keV} band, 
leaving the relative normalisations free to vary. 
We note, however, that some discrepancies between the pn and MOS seem to be 
present up to $\approx$~0.9~keV. 
The extension of our analysis down to 0.3~keV has the advantage of providing 
a slightly better constraint on the temperature of the soft thermal component without changing our main results.

\subsection{Spectral results: The 2000 June observation}

A single power-law model provides a poor fit to the \hbox{0.6--10~keV} pn$+$MOS data; 
strong residuals at low energies are evident in both instruments, as shown in Fig.~1. 
%
\begin{figure}
\includegraphics[angle=-90,width=85mm]{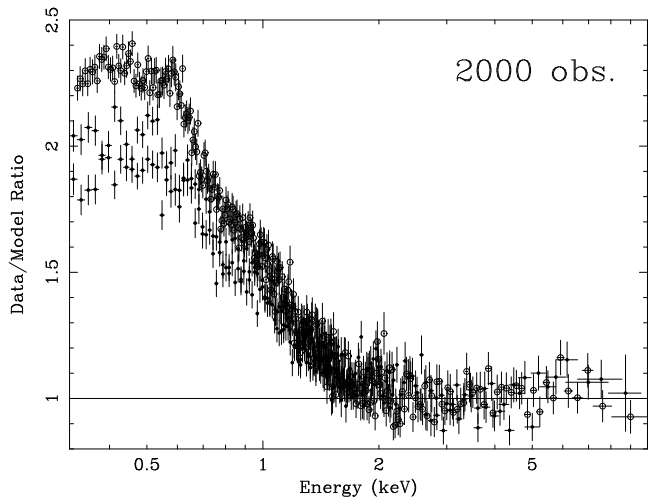}
\caption{pn (open circles) and MOS (filled circles) 
spectra fitted above 2~keV with a power law with $\Gamma\approx$~2.5 that has 
been extrapolated back to lower energies. An excess of counts is clearly seen 
below $\approx$~1.5~keV in all EPIC instruments.} 
\label{fig:ratio_onlypowerlaw_epic2000}
\end{figure}
%
The soft excess of Ark~564 rises sharply below $\approx$~1.5~keV 
similarly to the \sax\ observation (see Fig.~6 of Comastri et al. 2001). 
The presence of a soft component, parameterised either by a blackbody or a power law, 
is common across the NLS1 population (e.g., Comastri 2000; Vaughan et al. 2002). 
If thermal, this component is usually ascribed to the high-energy tail 
of the accretion disc responsible for the UV bump detected in the majority of NLS1s 
(see, e.g., Soria \& Puchnarewicz 2002 for discussion), although the rapid variability 
of the soft component detected from some NLS1s casts doubts on simple versions of the 
accretion-disc interpretation. 
Fitting the soft excess with a power law is inadequate; there is clear continuum 
spectral curvature that is not accounted for by a power law. 
Spectral curvature with similar shape has been detected in some other 
Seyfert galaxies (see, e.g., Pounds \& Reeves 2002 for a review; Collinge et al. 2001) 
and also in the narrow emission-line quasar PG~1211$+$143 (Pounds et al. 2003). 
The addition of a blackbody component is preferred by the EPIC data and finds support from 
previous \sax\ results (Comastri et al. 2001). 
This component has $kT=138^{+2}_{-3}$~eV [see model (1) in Table~2]. 
%
\begin{table}
\centering
\begin{minipage}{180mm}
\caption{2000 June observation: EPIC pn$+$MOS spectral results.}
\begin{tabular}{cccccc}
\hline
Model & $kT$ & $\Gamma$ & {\it E}$_{\rm edge}$ & $\tau$ & $\chi^{2}/d.o.f.$ \\
      & (eV) &          & (eV)           &        &                   \\
\hline
(1) & 138$^{+2}_{-3}$ & 2.55$\pm{0.02}$ & \dotfill & \dotfill & 1244/902 \\
\hline
(2) & 148$^{+2}_{-3}$ & 2.52$\pm{0.02}$ & 731$\pm{11}$ & 0.38$^{+0.07}_{-0.06}$ & 1081/900 \\
\hline
\end{tabular}
\end{minipage}
\begin{minipage}{180mm}
The edge threshold energy is in the source rest frame. 
\end{minipage}
\label{tab:tab2}
\end{table}
%

%
At higher energies, a power law with photon index $\Gamma=2.55\pm{0.02}$ dominates the spectrum. 
This model leaves significant residuals at soft \xray\ energies. 
The shape of these residuals (see Fig.~2) indicates a count deficit at $\approx$~0.7~keV. 
%
\begin{figure}
\includegraphics[angle=-90,width=85mm]{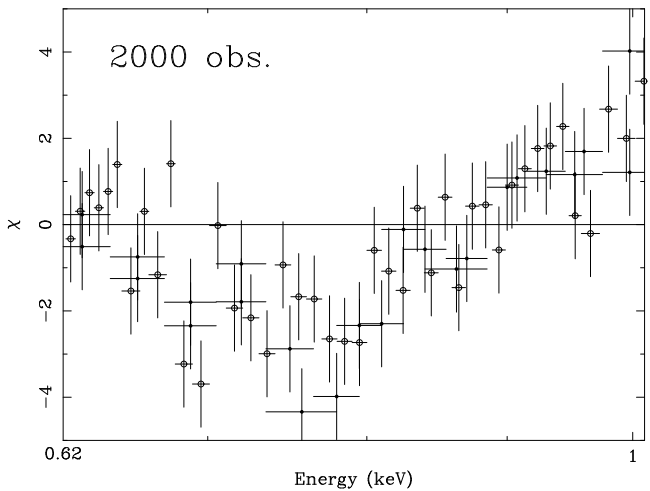}
\caption{Soft \xray\ data-to-model residuals (in terms of $\sigma$ with error 
bars of size one) obtained when the pn$+$MOS data 
are fitted with a blackbody plus power-law model [model (1) in Table~2]. 
Symbols are the same as in Fig.~1.}
\label{fig:delchi-edge_epic2000}
\end{figure}
%
We therefore added an absorption edge to the previous model [see model (2) in Table~2]. 
The edge provides a highly significant improvement in the fit quality 
($\Delta\chi^{2}$/$\Delta$d.o.f.=163/2). 
Its rest-frame threshold energy, 731$\pm{11}$~eV, is consistent with the edge being due to \ovii\ 
(which has an energy of 739~eV). 
The 68, 90, and 99 per cent confidence contours for the edge energy  
vs. optical depth (in the observed frame) are shown in Fig.~3. 
The edge is still required when other continuum models for the soft component (e.g., a power law, 
a double blackbody) are chosen, although the fit is significantly worse. 

The edge is also present in the \chandra\ High-Energy Transmission Grating (HETG) 
data taken simultaneously with the \xmm\ observation 
(see the spectrum shown in Fig.~3 of Matsumoto et al. 2001). 
The edge rest-frame threshold energy in the \chandra\ observation 
(712$^{+3}_{-4}$~eV; Matsumoto et al. 2001; Marshall 2002) 
is not in agreement with the nominal energy of the \ovii\ \hbox{K-edge}. 
The hypothesis that \chandra\ observed a highly redshifted \ovii\ \hbox{K-edge} 
with a velocity of \hbox{$\approx$~1000~km~s$^{-1}$} appears unlikely since the \ovii\ 
absorption lines detected by \chandra\ have no measurable velocity shifts (Matsumoto et al. 2001).  
We have performed an analysis of the RGS data searching for the spectral
features reported by Matsumoto et al. (2001) based upon HETG data.
However, we do not find any clear absorption lines 
(at most \hbox{$\Delta\chi^{2}$/$\Delta$d.o.f.=10/2} for a narrow absorption line) 
despite the high counting statistics of the RGS data. This is presumably due to the lower
spectral resolution of the RGS; 
the HETG spectral features are unresolved even with the higher spectral resolution of the HETG. 
%
%
Therefore, below we use the RGS mainly
to check the results derived from the EPIC spectral fitting rather than
presenting a detailed analysis of the RGS data.
%
\begin{figure}
\includegraphics[angle=-90,width=85mm]{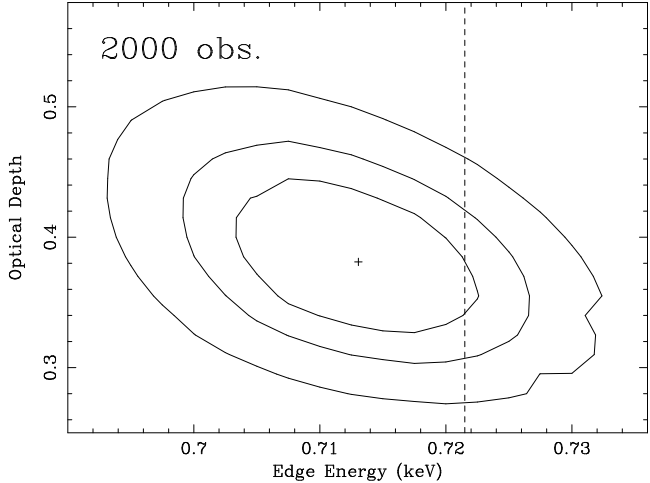}
\caption{68, 90, and 99 per cent confidence contours 
for the edge energy (in the observed frame) vs. optical depth. 
The dashed vertical line shows the redshifted energy of the \ovii\ K-edge.}
\label{fig:edge-tau_06210_epic2000}
\end{figure}
%
The edge is also present in the RGS spectra, although at a much lower significance level; 
spectral simulations suggest that this result is due to the combination of lower 
statistics and the presence of chip gaps in the RGS.  
The underlying continuum measured by the HETG 
is well parameterised by a model similar to the one used for our \xmm\ observation, 
even though the HETG bandpass is significantly narrower than the \xmm\ one. 
While the HETG best-fitting photon index ($\Gamma=2.56\pm{0.06}$) is consistent with the one 
measured by \xmm, the blackbody component has a lower temperature ($kT=124\pm{3}$~eV). 
The different blackbody temperature may be the cause of the different edge energy in the 
HETG observation. 

Adopting the blackbody plus power-law model and adding a 
warm absorber ({\sc absori} in {\sc xspec}; Magdziarz \& Zdziarski 1995) 
to fit the absorption feature of Ark~564 does not provide good results; 
significant residuals are still present at the energy corresponding to the \ovii\ edge. 
The ionisation parameter, defined as $\xi=\frac{L}{nR^2}$, 
where $L$ is the source luminosity, $n$ the warm absorber density, and $R$ the distance of the 
absorber from the photoionising source, is 73$^{+43}_{-35}$~erg~cm~s$^{-1}$. 
For reasonable assumptions of the physical parameters, 
the strongest bound-free soft \xray\ absorption features in the warm absorber model are the \ovii\ and \oviii\ K-edges 
(e.g., Reynolds \& Fabian 1995; Reynolds 1997; George et al. 1998). 
The relatively large optical depth of the detected \ovii\ edge 
probably causes the failure of the warm absorber model, since 
{\sc absori} also predicts the presence of the \oviii\ edge which is not observed in the data 
($\tau$$<$0.08). 
Although data with better resolution and higher signal-to-noise ratio 
are required to provide a physical justification for the presence of 
the \ovii\ edge and the absence or weakness of the \oviii\ edge, 
we note that similar 
spectral results have been found for the Seyfert~1 galaxy H~1419$+$480 using \xmm\ data 
(Barcons, Carrera \& Ceballos 2003). 
%
Although the different spectral resolution does not allow a proper comparison of the 
absorption features at different wavelengths, there are claims that the 
UV and X-ray absorbers in Ark~564 are physically related 
and possibly identical (e.g., Romano et al. 2002; Crenshaw et al. 1999, 2002; 
Matsumoto, Leighly \& Marshall 2002). 
However, the $N_{\ovii}$ value we derive from the observed K-edge optical depth 
($\approx1.4\times10^{18}$~cm$^{-2}$) is about an order of magnitude larger than 
the value predicted in Table~4 of Crenshaw et al. (2002) using photoionisation models and 
assuming the UV absorber to be a single zone. 
It is likely that a more complex structure for the absorber is 
required to reproduce both the UV and \xray\ data. 

%
%

In 2000 June Ark~564 was also observed by \asca\ (T01) 
with a 35-day observation (from 2000 June 1 to July 6). 
Unfortunately, the lower statistics in the portion of \asca\ data 
(re-analysed using standard procedures; $\approx$~12000 counts per instrument) 
taken simultaneously with our \xmm\ observation 
and the exclusion of the data below 0.7~keV 
do not allow one to reveal the \ovii\ edge detected by \xmm.\footnote{See $\S$2 of 
T01 for a discussion of the \asca\ calibration problems that led 
the authors to avoid the spectral band below 0.7~keV.} 
However, the spectral parameters for the blackbody \hbox{($kT=146\pm{9}$~eV)} 
and the power-law (\hbox{$\Gamma=2.46\pm{0.03}$}) components are in relatively 
good agreement with our results. 

At high \xray\ energies, 
the long \asca\ observation indicates that the 
iron K$\alpha$ emission line is complex, best fitted using a relativistic 
iron line produced by an accretion disc in either the 
Schwarzschild or Kerr metrics ({\sc xspec} models 
{\sc diskline} and {\sc laor}, respectively; Fabian et al. 1989; Laor 1991), 
and variable on time scales of a few days. 
The analysis of the \asca\ data exactly simultaneous with our \xmm\ observation, 
however, indicates only marginal evidence ($\Delta\chi^{2}$/$\Delta$d.o.f.=6/2) 
for an ionised \feka\ line (EW=113$^{+181}_{-79}$~eV). 
This is possibly caused by the relatively poor statistics in the short time interval 
simultaneous with the \xmm\ observation. 
%
%
\begin{figure}
\includegraphics[angle=-90,width=85mm]{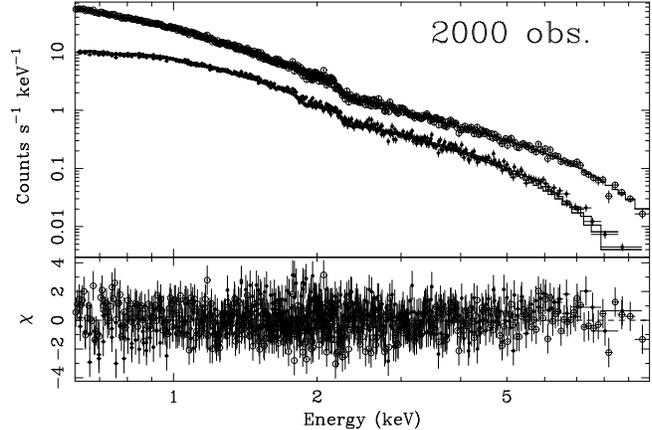}
\caption{Best-fitting spectrum [model (2) in Table~2] for the 
EPIC 2000 June data. The data-to-model residuals are shown in the bottom 
panel in units of $\sigma$. Symbols are the same as in Fig.~1.}
\label{fig:spectrum-bestfit_epic2000}
\end{figure}
%
The addition of a narrow (\hbox{$\sigma=10$~eV}) emission line improves the EPIC fit quality by 
$\Delta\chi^{2}$=8 at most when all of the line parameters, except for its normalisation, are frozen. 
We obtain EW=32$^{+46}_{-17}$~eV, EW=35$^{+30}_{-31}$~eV, and EW=43$^{+36}_{-35}$~eV for the neutral 
(6.40~keV), He-like (6.70~keV), and H-like (6.97~keV) \feka\ lines, respectively. 
%
%
%
We tried to account for the small residuals at high energies, having a ``bump-like'' shape (see Fig.~4), 
using a relativistic iron line produced by an accretion disc. 
%
The {\sc diskline} model improves the fit by 
$\Delta\chi^{2}$/$\Delta$d.o.f.=12/2 when all the line parameters are frozen 
to the default values\footnote{We assumed the line originates 
between 10 and 1000 gravitational radii and adopted 
an emissivity law $r^{-q}$ with $q=2$ for the illumination pattern of the accretion disc; 
the inclination angle of the disc was fixed to 30$\degr$.}
excluding the line energy and its normalisation, which are left free to vary. 
The line rest-frame energy is 6.10$^{+0.31}_{-0.17}$~keV, while its intensity is 
EW=98$^{+48}_{-46}$~eV. 
Both the line energy and EW are lower than the values found by T01. 
This is probably due to the fact that this model fails in reproducing the peak of the residuals 
at $\approx$~7~keV in the present \xmm\ data. 
%
Using the {\sc laor} model\footnote{We fixed the innermost radius 
as the last stable orbit for a Kerr black hole and the outer radius 
at the maximum value allowed by the model (400 gravitational radii). 
We fixed the emissivity index to 3 and the inclination angle of the disk to 30$\degr$.}
the improvement is slightly more 
significant ($\Delta\chi^{2}$/$\Delta$d.o.f.=18/2). 
The line rest-frame energy, 
considerably higher (7.07$^{+0.27}_{-0.20}$~keV) than in the previous model, and 
its EW (471$^{+175}_{-86}$~eV) are consistent with the best-fit model 
of T01, where the line peaks at $\approx$~7~keV. 
However, in contrast with the \asca\ results shown by T01, 
the presence of a complex \feka\ line is not strongly motivated by the present \xmm\ data, 
either by the residuals to a power-law fit or by the improvement in the $\chi^{2}$ after the inclusion of 
this feature. 
It is possible that residual uncertainties either in the calibration of \xmm\ spectra at high energies 
in ``Small-Window'' mode or in \asca, 
coupled with a rapid change in the intensity and shape of the line (already probed by the 
longer \asca\ observation), 
are responsible for the different results we obtained. 
Further investigations on this issue are required. 
%

When a reflection component is added to the blackbody plus power-law model, a much steeper 
photon index is obtained (as expected, given the known correlation of these parameters in 
the reflection models available in {\sc xspec}; 
see Vaughan \& Edelson 2001 and Perola et al. 2002 for discussion), but the fit does not 
improve significantly. The generally flatter photon index in MOS1 data (see Molendi \& Sembay 2003) 
and, possibly, problems in the pn-MOS cross-calibration using the ``Small-Window'' mode 
could produce a spurious reflection component in the MOS data. The significantly higher statistics 
collected by the EPIC pn at high energies 
reveal no evident additional spectral components and indicate that 
model (2) in Table~2 is a good parameterisation of the Ark~564 spectrum in the 2000 June observation. 
However, motivated by the discoveries of sharp drops in the \xmm\ spectra of 
two other NLS1s, 1H~0707$-$495 (Boller et al. 2002) and IRAS~13224$-$3809 (Boller et al. 2003), 
we fitted the 2--10~keV data using the best-fitting power law obtained in the 2--7~keV band (thus 
avoiding any problems with the \ovii\ edge and the soft thermal component). 
In contrast with the results obtained for the other two NLS1s, we find only marginal evidence for 
a drop at \hbox{$\approx$~7--8~keV} ($\Delta\chi^{2}$/$\Delta$d.o.f.=6/2 when the drop is 
parameterised by an absorption edge). 
This is a further indication that the present \xmm\ data do not require 
more complex modeling at high energies. 
%
%

\subsection{Spectral results: The 2001 June observation}

The 2001 June \xmm\ EPIC observation of Ark~564 is characterized by lower statistics (see Table~1) than 
the 2000 June observation. The exposure time, after the screening procedure, is relatively low, and the 
source has a lower flux level (see $\S$4). 
Following $\S$3.1, a blackbody plus power law is used to model the broad-band \xray\ continuum of 
Ark~564; the blackbody temperature and the power-law photon index [see model (1) in Table~3] 
are consistent with the corresponding model in the 2000 June observation (see Table~2). 
The poor quality of the fit, however, is suggestive of additional spectral complexity, 
in particular below $\approx$~1~keV. 
%
%
\begin{table}
\centering
\begin{minipage}{180mm}
\caption{2001 June observation: EPIC pn$+$MOS spectral results.}
\begin{tabular}{cccccc}
\hline
Model & $kT$ & $\Gamma$ & {\it E}$_{\rm edge}$ & $\tau$ & $\chi^{2}/d.o.f.$ \\
      & (eV) &          & (eV)           &        &                   \\
\hline
(1) & 136$\pm{3}$ & 2.56$\pm{0.03}$ & \dotfill & \dotfill & 968/757 \\
\hline
(2) & 144$^{+4}_{-3}$ & 2.54$\pm{0.03}$ & 727$^{+15}_{-16}$ & 
0.45$^{+0.13}_{-0.06}$ & 900/755 \\
\hline
\end{tabular}
\end{minipage}
\begin{minipage}{180mm}
The edge threshold energy is in the source rest frame.
\end{minipage}
\label{tab:tab3}
\end{table}
%

%
%
\begin{figure}
\includegraphics[angle=-90,width=85mm]{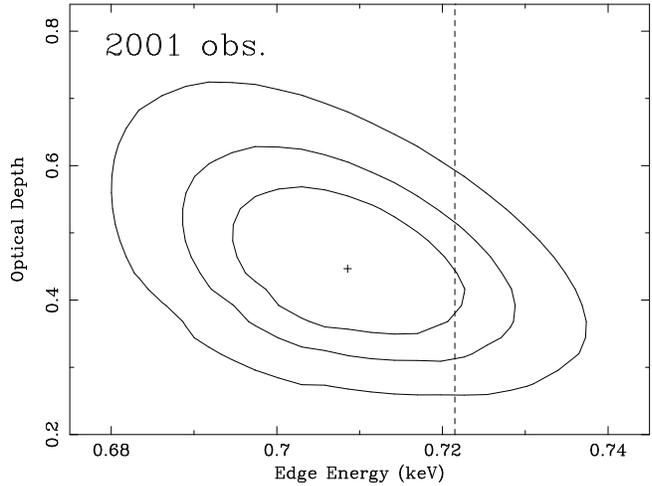}
\caption{68, 90, and 99 per cent confidence contours 
for the edge energy (in the observed frame) vs. optical depth. 
The dashed vertical line shows the redshifted energy of the \ovii\ K-edge.}
\label{fig:zbb_po_pn2001}
\end{figure}
%
Based on the 2000 observation and the shape of the residuals in the 2001 observation, 
we tried to account for the poor spectral fit by adding an absorption edge 
[model (2) in Table~3].  
%
%
\begin{figure*}
\includegraphics[angle=-90,width=85mm]{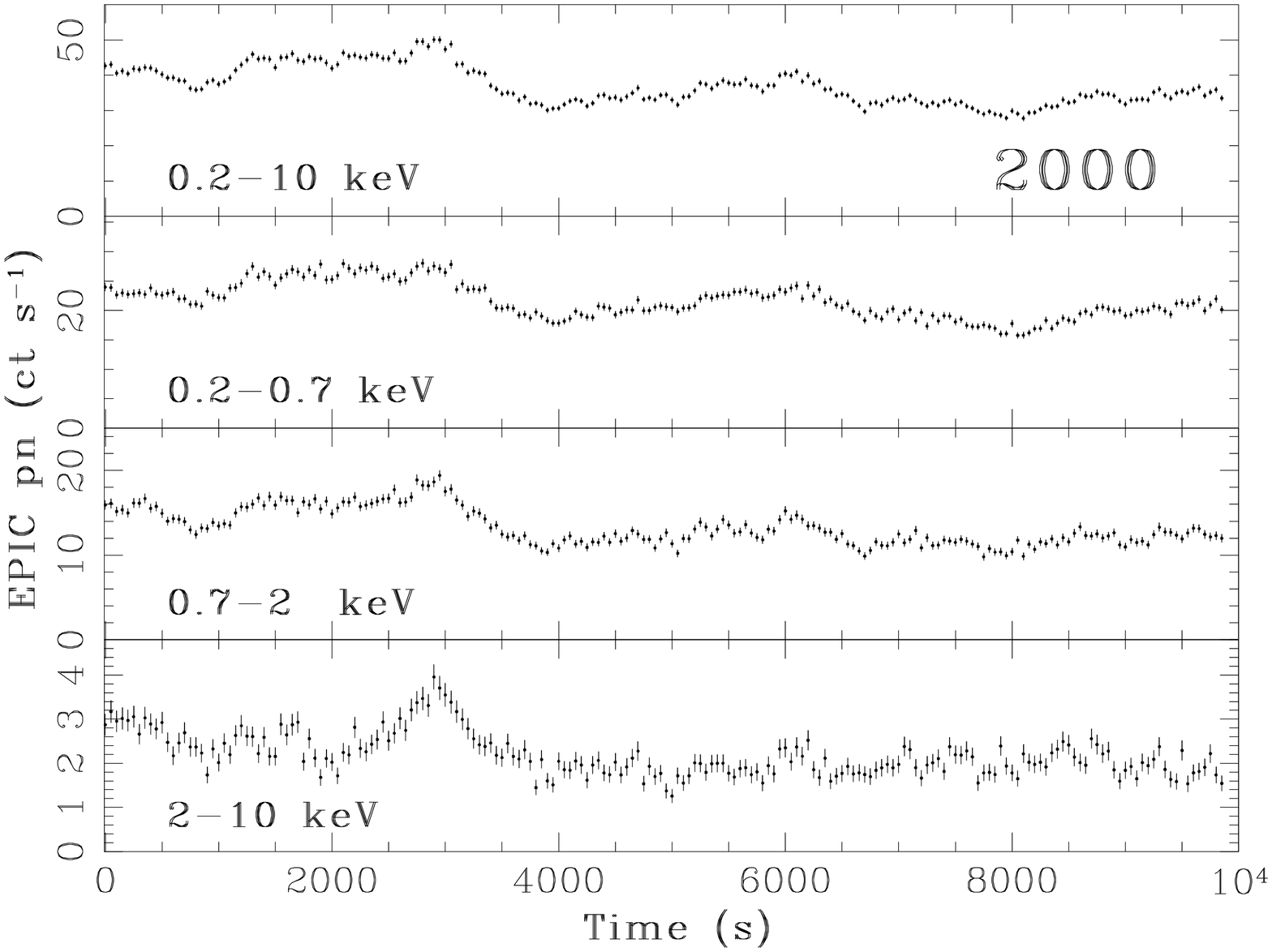}
\includegraphics[angle=-90,width=85mm]{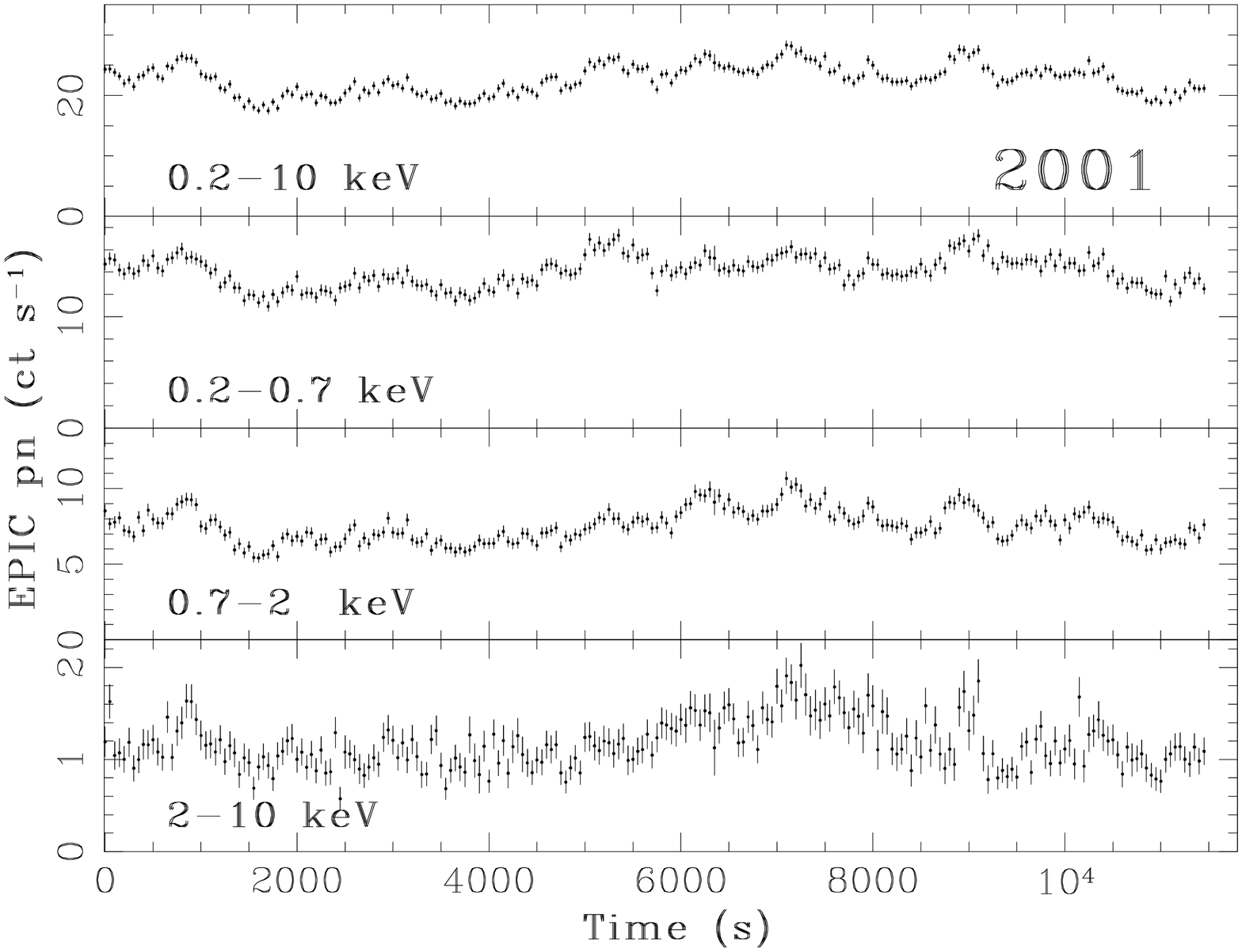}
\caption{2000 (left panel) and 2001 (right panel) pn light curves (50-s time bins). 
From top to bottom: \hbox{0.2--10~keV}, \hbox{0.2--0.7~keV}, \hbox{0.7--2~keV}, and \hbox{2--10~keV}. 
The count rate has not been scaled 
for detector efficiency ($\approx$~71\% in ``Small-Window'' mode).}
\label{fig:2k_lc}
\end{figure*}
%
The edge is significant statistically ($\Delta\chi^{2}$/$\Delta$d.o.f.=68/2); 
its rest-frame threshold energy, $727^{+15}_{-16}$~eV, is consistent with the 
\ovii\ \hbox{K-edge} (see Fig.~5), as in the 2000 June observation, and 
also the edge optical depth agrees with the previous EPIC observation. 
Some indications of this spectral feature are also present in the RGS data. 
There is no clear evidence for additional spectral features in the 2001 \xmm\ data. 
An upper limit of $\tau=0.12$ is derived for the \oviii\ optical depth; adopting a warm absorber model 
to reproduce the \ovii\ edge provides a worse spectral fit 
and a poorly constrained ionisation parameter. 

%
Although in the 2001 June observation of Ark~564 the hard \xray\ photon statistics are limited, 
we can place some constraints on the presence of \feka\ emission lines (both neutral and ionised). 
We find marginal evidence ($\Delta\chi^{2}$/$\Delta$d.o.f.=12/2; 
the width of the line is frozen to \hbox{$\sigma=10$~eV}) for a Gaussian \feka\ line. 
The line rest-frame energy ($6.16^{+0.05}_{-0.02}$~keV), 
lower than that expected for the neutral \feka\ emission line, coupled with its low significance 
casts some doubts on the nature of this feature. 
%
%
If we freeze the line energy at 6.40~keV, 
6.70~keV, and 6.97~keV for the neutral, He-like, and H-like \feka\ lines, 
we obtain EW$<71$~eV, EW=70$^{+60}_{-61}$~eV, and EW$<114$~eV, respectively, 
but none of these features is significant statistically. 
The absence or weakness of iron lines is supported by the lack of a strong reflection component, 
as the relatively good fit obtained at high energies with model (2) also suggests.

\section{Flux variability}

Between the 2000 and 2001 observations, Ark~564 shows significant flux variability. 
In the 2000 June observation, the observed soft (\fs) and hard (\fh) band 
fluxes are \hbox{(4.7--5.4)$\times10^{-11}$~\cgs} and 
\hbox{(2.1--2.4)$\times10^{-11}$~\cgs}, respectively, depending on the EPIC camera used 
and adopting the best-fitting models reported in Table~2. 
These correspond to Galactic absorption-corrected luminosities 
of \hbox{(8.8--10.0)$\times10^{43}$~\lum} and 
\hbox{(3.0--3.4)$\times10^{43}$~\lum}, respectively. 
In 2001 June the source shows a flux level lower by \hbox{$\approx$~40--50} per cent. 
We found \hbox{\fs=(3.0--3.2)$\times10^{-11}$~\cgs} and 
\hbox{\fh=(1.1--1.2)$\times10^{-11}$~\cgs} [model (2) in Table~3], corresponding 
to intrinsic luminosities of \hbox{(5.7--6.0)$\times10^{43}$~\lum} in the soft band and 
\hbox{(1.7--1.8)$\times10^{43}$~\lum} in the hard band. 
In both observations, the presence of a larger luminosity in the soft band than in the hard band 
argues against a simple model where the soft excess is just due to reprocessed 
hard \xray\ emission. 

For each observation, 
pn light curves were extracted in the \hbox{0.2--10~keV} range for 
continuous periods of 9.9 and 11.5 ks from the 2000 and 2001 data sets, 
respectively, and binned to 50-s time resolution (see Fig.~6, top panels).\footnote{We used 
the total (whole interval) 
light curves irrespective of the periods of ``flaring background'' because the source 
count rate is much higher than the background even during the flares. 
The background flares are weak compared to many seen by \xmm, and the 
exclusion of these intervals for the spectral analysis was conservative. 
Their effect on the PSD derived in $\S$5 is negligible.
Similarly, the exclusion of the \hbox{0.2--0.3~keV} data from the spectral analysis 
was a conservative choice.}
The light curves were background subtracted and exposure corrected 
(i.e., corrected for telemetry drop outs). 
Analyses of the pn light curves 
in different energy bands 
(\hbox{0.2--10~keV}, \hbox{0.2--0.7~keV}, \hbox{0.7--2~keV}, and \hbox{2--10~keV}; see Fig.~6) 
show that the source is highly variable, 
both at low and high energies. The rapid variability of the soft component 
and its relatively high temperature (see $\S$3.1 and 3.2) suggest that it is 
unlikely to be simple thermal radiation from an accretion disc. 
It is possible that it is quasi-thermal emission originally from the 
accretion disc that has been upscattered in a hot surface layer of the disc or the corona. 
Alternatively, the soft component might be due to turbulent Comptonisation 
occurring close to the black hole (Socrates, Davis \& Blaes 2003).

\section{Power spectral density function}

The high-frequency power spectral density function (PSD) of Ark~564 was 
estimated from the pn light curves 
using the method of Vaughan, Fabian \& Nandra (2003). 
%
pn 2000 and 2001 periodograms were computed using standard Fourier methods. 
The two periodograms appeared consistent with one another, and therefore a 
single binned periodogram was produced by combining the two periodograms 
and binning, using the method of Papadakis \& Lawrence (1993), such that 
there are 15 periodogram estimates per frequency bin. Figure~7 shows the 
average periodogram. At high frequencies the power is dominated by the 
Poisson noise. 
%
%
\begin{figure}
\includegraphics[angle=-90,width=85mm]{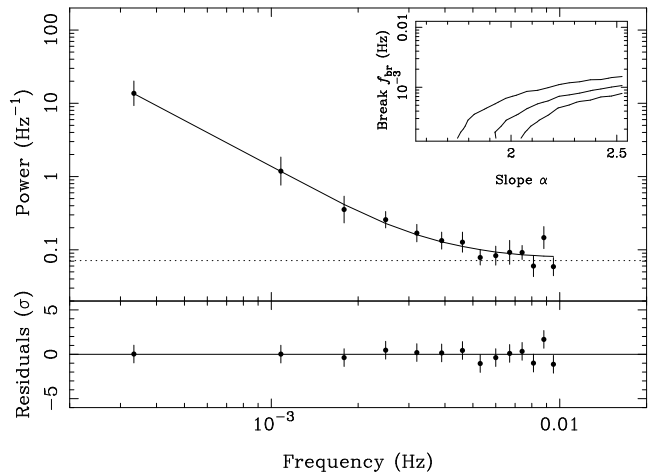}
\caption{(Upper panel) High-frequency periodograms estimated using the 0.2--10~keV light curves 
and 50-s resolution. The solid curve shows the best-fitting model; 
the background noise level (from Poisson noise) is indicated with the dotted line. 
The 68, 90, and 99 per cent confidence contours 
for the two interesting parameters (the power-law index and the 
break frequency) are shown in the insert. 
(Lower panel) Data-to-model residuals in units of $\sigma$.}  
\label{fig:psd}
\end{figure}
%
At low frequencies the source variations dominate with a 
steep, approximately power-law spectrum rising toward low frequencies. 

From previous monitoring with \rxte\ and \asca\ it is known that 
the low-frequency PSD is well represented by a broken power law, with a 
power-law index of $\approx 0$ below a frequency \hbox{$f_{\rm br} \approx 1.6 \times 10^{-6}$~Hz} 
and an index of $\approx 1.2$ above the break (Markowitz et al. 2003). 
The \xmm\ data at much higher frequencies show a much steeper spectrum and imply 
the PSD must show a high-frequency break as well, confirming the suggestion of 
Papadakis et al. (2002) on the basis of \asca\ data.  In order to estimate the high-frequency PSD, 
these data were fitted using the Monte Carlo procedure of Vaughan et al. (2003) 
with a doubly broken power-law model. The power spectral index 
was fixed at $0$ below $10^{-6}$ Hz and $1.2$ above this frequency. 
Then a second, high-frequency break was included in the model with both the 
break frequency ($f_{\rm br}$) and the spectral index above the break 
($\alpha$) as free parameters in the fitting.\footnote{The low-frequency PSD 
needed to be specified in the model as the fitting procedure accounts for 
biases in the periodogram, notably ``red-noise leak'', that depend on the 
PSD at low frequencies. See Vaughan et al. (2003) for details on this.}

This model provided a statistically acceptable fit to the data 
($\chi_{\nu}^2 < 1.0$), but the free parameters were poorly constrained. 
The PSD is steep, with an index $\alpha > 1.95$, but the location of the 
break to this steep slope is only constrained to be 
\hbox{$f_{\rm br} < 8 \times 10^{-4}$~Hz} (90 per cent confidence limit for one interesting parameter). 
This fit thus supports the claim by Papadakis et al. (2002) that Ark~564 
shows a doubly broken PSD, although the inferred position of the break 
is a factor $\approx$~2 lower than that obtained in their analysis. 

Curiously, the doubly broken PSD resembles that seen in Cyg X-1 in its low/hard 
state, contrary to the expectation that Ark~564 might look more like the 
high/soft state (which only shows the high-frequency break). 
However, it is vital to constrain the high-frequency PSD and to confirm the second break 
in the PSD before a fair comparison can be made between Ark~564 and Galactic black hole candidates 
(which can also show ``very high'' and ``intermediate'' state PSDs).

\section{Summary}

We have reported two \xmm\ observations of the bright NLS1 Ark~564 taken 
one year apart. 
The principal results are the following: 
\begin{itemize}
\item 
A blackbody with \hbox{$kT\approx$~140--150~eV} 
and a power law with \hbox{$\Gamma\approx$~2.50--2.55} provide a good parameterisation of the 
\xray\ continuum. 
No significant spectral changes are observed between the two observations, 
although the \xray\ flux in the second observation is lower by $\approx$~40--50 per cent. 
\item
In both observations a significant absorption edge at $\approx$~0.73~keV, corresponding to the 
\ovii\ K-edge, is detected. 
\item
We find tentative evidence for an iron emission line (fitted using a {\sc laor} model) 
in the 2000 June observation. 
\item
The present \xmm\ data support the idea that the PSD shows two breaks. 
A longer observation is required to place stronger constraints on the location 
of the high-frequency break. 
\end{itemize}

\section*{Acknowledgments}
The work reported herein is based on observations obtained with \xmm, 
an ESA science mission with instruments and contributions directly funded by ESA 
Member States and the USA (NASA). 
We gratefully acknowledge the financial support of NASA LTSA grant NAG5-13035 (CV, WNB) 
and NASA grant NAG5-9939 (CV, WNB). 
CV also acknowledges partial support from the Italian Space Agency under contract 
ASI I/R/073/01. 
ACF thanks the Royal Society for support. 
We thank D.~Alexander, A.~Comastri and G. Matt for interesting suggestions, 
S.~Molendi and P.~Ranalli for useful discussions about \xmm\ calibrations, 
L. Angeretti for help with the SM macros, 
and the referee for her/his thoughtful comments and suggestions which improved 
the quality of the paper.

\bsp

\label{lastpage}

\end{document}